\documentclass[nofootinbib,aps,prd,preprintnumbers,superscriptaddress,notitlepage,12pt]{revtex4-1}
\pdfoutput=1

\usepackage{amsmath,amssymb,graphicx}
\usepackage{multirow,enumerate,cancel,nicefrac,color,slashed}
\usepackage[colorlinks=true,citecolor=black,linkcolor=black,urlcolor=blue,hypertexnames=false]{hyperref}
\usepackage[margin=1.0in]{geometry}

\graphicspath{{figures/}}

\def\bea{\begin{eqnarray}}
\def\eea{\end{eqnarray}}

\usepackage{accents}
\newlength{\dhatheight}

\usepackage{color}

\usepackage[usenames,dvipsnames]{xcolor}

\begin{document}

\preprint{FERMILAB-PUB-15-551-T}
\preprint{UCI-HEP-TR-2015-20}

\title{Vector Dark Matter through a Radiative Higgs Portal}

\author{Anthony DiFranzo}
\affiliation{Department of Physics \& Astronomy, University of California, Irvine, CA USA}
\affiliation{Theoretical Physics Department, Fermilab, Batavia, IL USA}
\author{Patrick J. Fox}
\affiliation{Theoretical Physics Department, Fermilab, Batavia, IL USA}
\author{Tim M.P.\ Tait}
\affiliation{Department of Physics \& Astronomy, University of California, Irvine, CA USA}

\begin{abstract}
We study a model of spin-1 dark matter which interacts with the Standard Model predominantly via exchange of Higgs bosons. 
We propose an alternative UV completion to the usual Vector Dark Matter Higgs Portal, in which 
vector-like fermions charged under SU(2)$_W \times$ U(1)$_Y$ and under the dark gauge group, U(1)$^\prime$, 
generate an effective interaction between the Higgs and the dark matter at one loop. 
We explore the resulting phenomenology and show that this dark matter candidate is a viable thermal relic and satisfies Higgs invisible width constraints as well as direct detection bounds.
\end{abstract}

\maketitle


\section{Introduction}
As the only elementary scalar in the Standard Model (SM), the Higgs boson presents a unique opportunity as a window to physics beyond the Standard Model (SM).
The operator $H^\dagger H$ is the lowest dimensional operator which is both a gauge and Lorentz singlet.
As such, it occurs time and again as the means by which physics uncharged under the SM gauge symmetries communicates with the Standard Model.  In particular,
it is an effective mechanism by which scalar
dark matter (DM) can talk to the ordinary matter \cite{Burgess:2000yq}, as is required if we wish to understand its abundance in the Universe
today as the result of thermal processes acting in a standard cosmological history.

In the present work, we focus on the case in which the dark matter is a spin one vector boson.  At first glance, it would appear that this case
(much like scalar DM) offers a renormalizable connection between the dark matter and the Higgs \cite{Djouadi:2011aa,Lebedev:2011iq},
\begin{equation}
\mathcal{L} \supset \lambda ~H^\dagger H ~V_\mu V^\mu~,
\label{eq:naive}
\end{equation}
where $V_\mu$ is a massive vector field which plays the role of dark matter and $\lambda$ is a dimensionless coupling.
But this form, while invariant under the SM gauge symmetries, is misleading.  Just like the SM $W$ and $Z$ bosons, a well-behaved UV description of $V$
requires that it be associated with a gauge symmetry (the most simple construction of which would be an Abelian U(1)$^\prime$, though one could also consider
non-Abelian theories as well), spontaneously broken to give $V$ a mass.  The term in Eq.~(\ref{eq:naive}) violates the U(1)$^\prime$, and must be engineered
via its spontaneous breaking.

One tempting avenue would be to charge the Higgs itself under U(1)$^\prime$.  In that case the Higgs kinetic term $(D_\mu H)^\dagger (D^\mu H)$ contains
Eq.~(\ref{eq:naive}), and the mass of $V$ will arise as part of the vacuum expectation value (VEV) of $H$, naturally connecting the scale of the $V$ mass
to the electroweak scale.  However, this construction contains other terms which mix $V$ with the SM $Z$ boson, with the result that $V$ will inevitably
end up unstable and contribute unacceptably to precision electroweak measurements unless it is very light (implying that it is very weakly coupled).
This regime, though worth pursuing, is not very interesting for particle physics at the weak scale, and not very amenable to exploration through
Higgs measurements at the LHC.

The situation is very different when the $V$ mass is the result of a VEV living in a different scalar particle $\Phi$ which is a SM gauge singlet.  In that case,
there is no dangerous mixing with the SM $Z$ boson, and the gauge coupling can be relatively large,
\begin{eqnarray}
{\cal L} &~ \supset ~ & -\frac{1}{4} V_{\mu \nu} V^{\mu \nu} + \left( D_\mu \Phi \right)^\dagger \left( D^\mu \Phi \right) - V (\Phi)
+ \lambda_P ~ |H|^2 |\Phi|^2~,
\label{eq:module1}
\end{eqnarray}
where $D_\mu \Phi \equiv \partial_\mu \Phi - g Q_\Phi V_\mu \Phi$ is the usual covariant derivative for a particle of charge $Q_\Phi$
and $V(\Phi)$ is a U(1)$^\prime$-invariant potential designed to induce a VEV $\langle \Phi \rangle = v_\phi$, producing a mass for $V$,
\begin{eqnarray}
m_V^2 & = &  g^2Q_\Phi^2~ v_\phi^2~.
\label{eq:Vmass}
\end{eqnarray}
We have also included a scalar Higgs portal coupling $\lambda_P$, which leads to tree-level
mixing between the SM Higgs boson and the Higgs mode of $\Phi$, effectively implementing the Higgs portal.
As a construction implementing the Higgs portal, it is well motivated and has been extensively explored in the 
literature\footnote{It also provides a mechanism to stabilize the Higgs potential \cite{Duch:2015jta} 
and/or generate a first order electroweak phase transition \cite{Chao:2014ina}.}
\cite{Hambye:2008bq,Farzan:2012hh,Baek:2012se,Baek:2013qwa,Baek:2014jga,Baek:2014goa,Ko:2014gha,Gross:2015cwa,DiChiara:2015bua,Chen:2015dea,Kim:2015hda}. 

However, it does not represent the {\em only} possible UV completion.  In this work, we explore an alternative completion which realizes the Higgs portal
as a consequence of additional heavy fermions which are charged under both U(1)$^\prime$ and the SM gauge symmetries.  
At one loop, these fermions mediate an interaction between the Higgs and the DM somewhat in analogy with the effective Higgs-gluon vertex
induced by the top quarks in the SM.
This {\em radiative} UV completion leads to different phenomenology and singles out different interesting regions of parameter space.

This article is organized as follows.  In Sec.~\ref{model}, we discuss a simplified picture to illustrate the most important physics behind this concept, followed by the 
full matter content of the UV theory. In Sec.~\ref{results}, we examine the phenomenology in light of experimental probes, such as direct detection, the invisible Higgs width, 
and relic abundance. We first focus on the case where the simplified picture is valid, with and without also considering mixing generated by a Scalar Higgs Portal. 
We then examine the effect of the full radiative portion of the UV theory.
We reserve Sec.~\ref{conclusion} for conclusions and summary.

\section{Radiative Higgs Portal for Vector Dark Matter}
\label{model}
\subsection{Particle Content and Structure}

A radiative model often has multiple paths to the same low energy physics, since the mediating particles are not themselves involved in the initial and final states.
Starting with the basic module of Eq.~(\ref{eq:module1}), we aim for a construction which adds fermions mediating an interaction of the form~(\ref{eq:naive}) such
that:
\begin{itemize}
\item the vector particle $V$ remains stable at the radiative level, which in particular requires that it does not kinetically mix with the SM electroweak interaction;
\item the full gauge structure SU(3)$_C \times$ SU(2)$_W \times$ U(1)$_Y \times$ U(1)$^\prime$ remains free from gauge anomalies;
\item there are no large contributions to the SM Higgs coupling to gluons or photons in contradiction with LHC measurements \cite{Flechl:2015foa}.
\end{itemize}
The first of these is the most subtle.  Generically, communication between the SM Higgs and $V$ requires that the mediator fermions be charged under
both U(1)$^\prime$ and the Standard Model, which typically will induce processes involving an odd number of $V$'s, resulting in their decay.  The simplest example of
such a process is the kinetic mixing between $V$ and hypercharge.   Such dangerous processes can be forbidden by a
charge-conjugation symmetry, under which $V$ is odd.  In analogy with Furry's theorem of QED \cite{Furry:1937}, this symmetry forbids
processes involving an odd number of $V$'s at energies below the masses of the mediator fermions.

\begin{table}
\centering
\caption{Charge assignments for fermions $\psi$, $\chi$, and $n$ and complex scalar $\Phi$.}
\begin{tabular}{cccc}
  ~~Field~~~ & ~~~(SU(2)$_W$, U(1)$_Y$, U(1)$^\prime$)~~~~~~~~~~~~~
  & ~~Field~~~ & ~~~(SU(2)$_W$, U(1)$_Y$, U(1)$^\prime$)
    \\
  \hline \hline                   
  $\psi_{1\alpha}$ & (2, \nicefrac{1}{2}, ~1)  &
  $\psi_{2\alpha}$ & (2, \nicefrac{1}{2}, -1)   \\
  $\chi_{1\alpha}$ & (2, \nicefrac{-1}{2}, -1) &
  $\chi_{2\alpha}$ & (2, \nicefrac{-1}{2}, ~1)   \\
  $n_{1\alpha}$ & (1, ~0, ~-1) &
  $n_{2\alpha}$ & (1, ~0, ~~1)  \\
  \hline
  $\Phi$~ & (1, ~0, ~~$Q_\Phi$) \\
\end{tabular}
\label{tab:NPtran}
\end{table}

Cancelling gauge anomalies further suggests that the additional fermions appear in vector-like pairs under both the SM and U(1)$^\prime$ gauge symmetries,
whereas renormalizable coupling to the Higgs requires fields in SU(2)$_W$ representations of size $n$ and $n+1$ (and have hypercharges differing by $1/2$).
A minimal set of particles satisfying these conditions is shown in Table~\ref{tab:NPtran}, consisting of four SU(2)$_W$ doublets and two singlets.
(Different) pairs of the doublets are vector-like under both U(1)$_Y$ and U(1)$^\prime$, cancelling gauge anomalies, and
 a U(1)$^\prime$ charge conjugation is implemented by $f_1 \! \leftrightarrow \! f_2$ (where $f = \psi, \chi, n$).

We have left the U(1)$^\prime$ charge of $\Phi$ as a free non-zero parameter which controls the dark matter mass as per Eq.~(\ref{eq:Vmass}).
Choosing $Q_\Phi=\pm1$ would allow the $\Phi$ VEV to mix the SM lepton doublets with the new fermions, which would be strongly constrained
by precision measurements and ruin the U(1)$^\prime$ charge conjugation symmetry.  Choosing $Q_\Phi=\pm2$ would allow for Yukawa interactions of
$\Phi$ with pairs of the new fermions, which would complicate the analysis of their mass eigenstates.  
We will restrict ourselves to other values for $Q_\Phi$, which avoids these features, and serves simply to adjust the mass of $V$.  
It's worth pointing out that this implies that the lightest of the fermionic states is also stable, and will be present in the Universe to some
degree as a second component of dark matter.  However, provided its mass is much larger than $m_V$, fermion
anti-fermion pairs will annihilate efficiently into weak bosons and $V$'s, leaving it as a negligible fraction of the dark matter.

In 2-component Weyl notation, the Lagrangian contains mass terms and Yukawa interactions for the new fermions,
\begin{equation}
\begin{aligned}
\mathcal L &~ \supset -m ~ \epsilon^{ab}  \left( \psi_{1a} \chi_{1b} + \psi_{2a} \chi_{2b} \right) - m_n ~ n_1 n_2\\
          &- y_{\psi}~\epsilon^{ab}  \left( \psi_{1a} H_b n_1 + \psi_{2a} H_b n_2 \right) - y_{\chi} \left( \chi_{1} H^* n_2 + \chi_{2} H^* n_1 \right) + h.c.
\end{aligned}
\end{equation}
where $a$ and $b$ are SU(2)$_W$ indices, the SM Higgs  $H$ is defined to transform as a (2,~$\nicefrac{-1}{2}$,~0),
and spin indices have been suppressed.  
The U(1)$^\prime$ charge conjugation symmetry, $f_1 \! \leftrightarrow \! f_2$ is manifest.
After electroweak symmetry-breaking, the mass terms can be written as,
\begin{eqnarray}
\label{eq:mass}
\mathcal L_m &= - N^T M_n N' - E^T M_e E' + h.c.
\end{eqnarray}
where
\begin{eqnarray}
N &=
\begin{bmatrix}
  \psi_{1n}\\
  \chi_{2n}\\
  n_2
\end{bmatrix},
~~~~~N' =
\begin{bmatrix}
  \psi_{2n}\\
  \chi_{1n}\\
  n_1
\end{bmatrix},
~~~~~E =
\begin{bmatrix}
  \psi_{1e}\\
  \chi_{2e}\\
\end{bmatrix},
~~~~~E' =
\begin{bmatrix}
  \chi_{1e}\\
  \psi_{2e}\\
\end{bmatrix},
\end{eqnarray}
assemble collections of the electrically neutral ($N$ and $N^\prime$) and charged ($E$ and $E^\prime$) components of the fermions, and the mass matrices
are given by,
\begin{eqnarray}
M_n =
\begin{bmatrix}
    0    & -m  & -y_{\psi}v/\sqrt{2}  \\
    -m   &  0  & y_{\chi}v/\sqrt{2} \\
    -y_{\psi}v/\sqrt{2}  &  y_{\chi}v/\sqrt{2}  &  m_n
\end{bmatrix}, ~~~~~
M_e =
\begin{bmatrix}
    m  &  0 \\
    0  &  m
\end{bmatrix}.
\end{eqnarray}
In the mass basis, there are three electrically neutral and two charged Dirac fermions, all of which interact with the dark matter $V$ diagonally, since the states that mix all carry the same $U(1)'$ charge.
Their coupling to the SM Higgs will involve the mixing matrices which transform from the gauge to the mass basis.

Note that by construction the electrically charged fermions
receive no contributions from $\langle H \rangle$, implying that they do not interact with the Higgs boson and lead to no
one-loop correction to its effective coupling to photons.
Our choice to arrange $N$ such that they also receive no contributions from $\Phi$ implies that the fermions do not renormalize the usual
Higgs portal coupling $\lambda_P$ of Eq.~(\ref{eq:module1}) at one-loop (starting at two loops, there are contributions mediated by a mixture
of the fermions and $V$ itself).  In order to better extract the features of the radiative model, we self-consistently assume that $\lambda_P$ is small
enough to be subdominant in the majority of the remainder of this work.

\subsection{$\sigma_{\rm SI}$ and Higgs Invisible Width}

\begin{figure}
\centering
\includegraphics[width=0.35\textwidth]{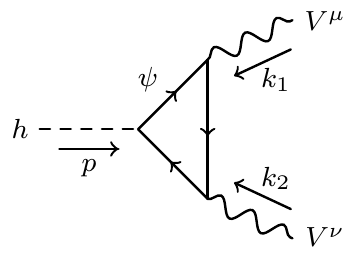}
\caption{Representative triangle diagram contributing to the Higgs--dark matter interaction.}
\label{fig:loop}
\end{figure}

Both the direct detection cross-section and the Higgs invisible decay width result from triangle diagrams (see Fig.~\ref{fig:loop}).  Integrating out the fermion $\psi$ running in the loop, the $h-V-V$ interaction can be encoded by two form factors:
\begin{equation}
 - \left( \frac{1}{4} A(p^2) ~h ~V^{\mu\nu}V_{\mu\nu} + \frac{1}{2} B(p^2)~ h ~V^{\mu}V_{\mu} \right)
\label{eq:eftloop}
\end{equation}
with coefficients $A$ and $B$ which are (in the on-shell DM limit, $k_1^2 = k_2^2 = m_V^2$) functions of the fermion masses and mixings, $m_V$, 
and the momentum through the Higgs line, $p^2$.  Reasonably compact analytic expressions for $A$ and $B$ are derived in Appendix~\ref{app:tri}.
We observe that $B(p^2)\rightarrow 0$ in the limit $m_V \rightarrow 0$ (i.e.\ when the $U(1)^\prime$ symmetry is restored), as is required by
gauge invariance, see Appendix~\ref{app:tri}. 

In terms of $A$ and $B$, the cross section for non-relativistic scattering of $V$ with a nucleon $n$ is given by,
\begin{eqnarray}
\sigma_{\rm SI} &= &\frac{1}{4\pi m_h^4} \left(\frac{f_n}{v}\right)^2 \left(\frac{m_n^2}{m_n+m_V}\right)^2 |B(0) - A(0) ~m_V^2|^2 \label{eq:loopsigma}
\end{eqnarray}
where the momentum transfer through the Higgs is approximated as $p^2 \approx 0$,
\begin{eqnarray}
f_n = \sum_{q=u,d,s} f_{Tq}^{(n)} + \frac{2}{9} f_{TG}^{(n)},
\end{eqnarray}
and we use the hadronic matrix elements $f_{Tq}$, from DarkSUSY \cite{Gondolo:2004sc}.
Because of the tiny up and down Yukawa
couplings, scattering mediated by a Higgs is to good approximation iso-symmetric.

The same three point vertex function also describes the invisible decay width of the Higgs boson,
\begin{eqnarray}
\Gamma(h \rightarrow V V) &=& 
\frac{1}{64\pi m_h} \sqrt{1-4\frac{m_V^2}{m_h^2}} \left[ \left| A(m_h^2) \right|^2 m_h^4 \left(1 - 4\frac{m_V^2}{m_h^2} + 6\frac{m_V^4}{m_h^4}\right) 
\right. \label{eq:loopwidth}\\ 
& &  \left. 
+ 6 \operatorname{Re}\left( A^*(m_h^2) B(m_h^2) \right) m_h^2 \left(1-2\frac{m_V^2}{m_h^2}\right) 
+ \frac{1}{2} \left| B(m_h^2) \right|^2 \frac{m_h^4}{m_V^4} \left(1 - 4 \frac{m_V^2}{m_h^2} + 12 \frac{m_V^4}{m_h^4} \right)\right] \nonumber
\end{eqnarray}
where the Higgs is on-shell, $p^2 = m_h^2$.  Note that because for small $m_V$ the coefficient $B(p^2) \propto m_V^4$, this expression is finite in
the limit $m_V \rightarrow 0$, as it should be.

\subsection{Annihilation Cross Section and Relic Abundance}

\begin{figure}
	\includegraphics[width=0.35\textwidth]{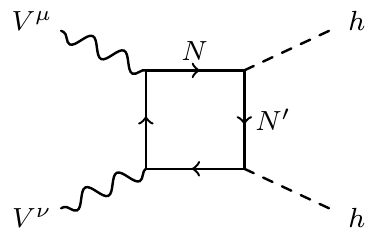}	\hspace*{1cm}
	\includegraphics[width=0.35\textwidth]{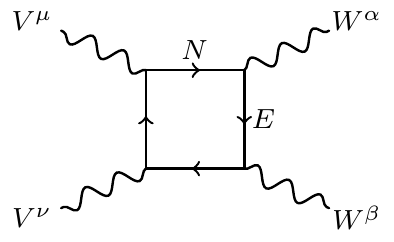}
	\caption{Representative box diagrams which contribute to DM annihilation into pairs of Higgs or electroweak bosons.}
\label{fig:box}
\end{figure}

Pairs of dark matter can annihilate through the three point coupling of Fig.~\ref{fig:loop} through an (off- or on-shell) SM Higgs, leading to final states containing heavy quarks
and/or weak bosons.  These contributions exhibit a strong resonant behavior when $m_V \simeq m_h / 2$.
The gauge and Higgs boson final states also receive contributions at the same order from box diagrams (see Fig.~\ref{fig:box}),
which contribute to processes including $VV \rightarrow hh, ZZ, WW, \gamma\gamma, hZ, Z\gamma$. These box
diagrams are sensitive to more of the details of the UV theory, receiving contributions from the charged fermions as well as the neutral ones.  As a result, simple
analytic forms are not particularly illuminating, and we evaluate them using FeynArts \cite{Hahn:2000kx}, FormCalc, and LoopTools \cite{Hahn:1998yk}.
In the following section, we compute the full annihilation cross section including all of the accessible SM final states.

\section{Experimental Constraints and Parameter Space}
\label{results}

In this section, we examine the interesting parameter space, finding the regions consistent with
the LUX limits on the spin independent DM-nucleon scattering cross-section \cite{Akerib:2013tjd}; and the invisible decay width
of the Higgs produced via vector boson fusion (VBF) as constrained by CMS with 19.7 fb$^{-1}$ at 8 TeV \cite{Chatrchyan:2014tja}. 
In the latter, we include the off-shell Higgs contribution following the technique presented in \cite{Endo:2014cca}, simulating VBF Higgs production with 
HAWKv2.0 \cite{Denner:2011id}.  We also identify the regions leading to the correct thermal relic abundance for a standard cosmology, computing
the loop diagrams with FeynArts \cite{Hahn:2000kx}, FormCalc, and LoopTools \cite{Hahn:1998yk}, which is then linked into micrOMEGAsV4.0 \cite{Belanger:2013oya}.

Because of the relatively large number of parameters, we build up insight into the phenomenology gradually
by considering three different limits of the full theory.  Initially
in Sec.~\ref{sec:singleF},
we consider the limit in which one of the neutral fermions is much lighter than both the other two neutral states and both of the charged ones, and the 
coupling $\lambda_P$ is small enough to be neglected.  We follow this in Sec.~\ref{sec:singleHP} by allowing $\lambda_P$ to be large enough that there
is relevant mixing between $h$ and the Higgs mode of $\Phi$.  Finally, in Sec.~\ref{sec:fullF} we switch off $\lambda_P$ once more, but consider the case
where all mediator fermions have comparable masses.

\subsection{Single Fermion Limit}
\label{sec:singleF}

We begin with the case where the charged fermions and the two heavier neutral states are much heavier than the lightest neutral state, effectively
decoupling from the phenomenology, and $\lambda_P$ can be ignored.  As before we assume the physical scalar contained in $\Phi$ is heavy enough to be ignored.  In this limit, the relevant 
parameters are the $U(1)^\prime$ gauge coupling $g$, Yukawa coupling to the light fermion $y$, light fermion mass $m_\psi$, and the vector dark matter mass $m_V$. 
As we will see below, the correct thermal relic density can only be achieved for annihilation in the Higgs funnel region, for which one can neglect the box diagram contributions.
In that case, the gauge and Yukawa couplings always appear in the combination $y g^2$, leaving only three relevant parameter combinations.

\begin{figure}
	\includegraphics[width=0.475\textwidth]{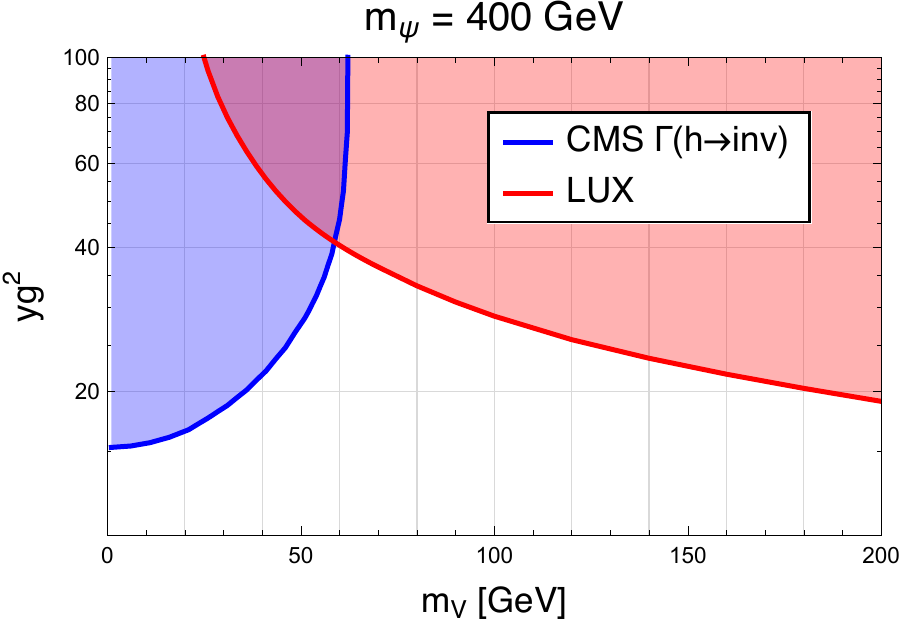} \hspace*{0.5cm}
	\includegraphics[width=0.475\textwidth]{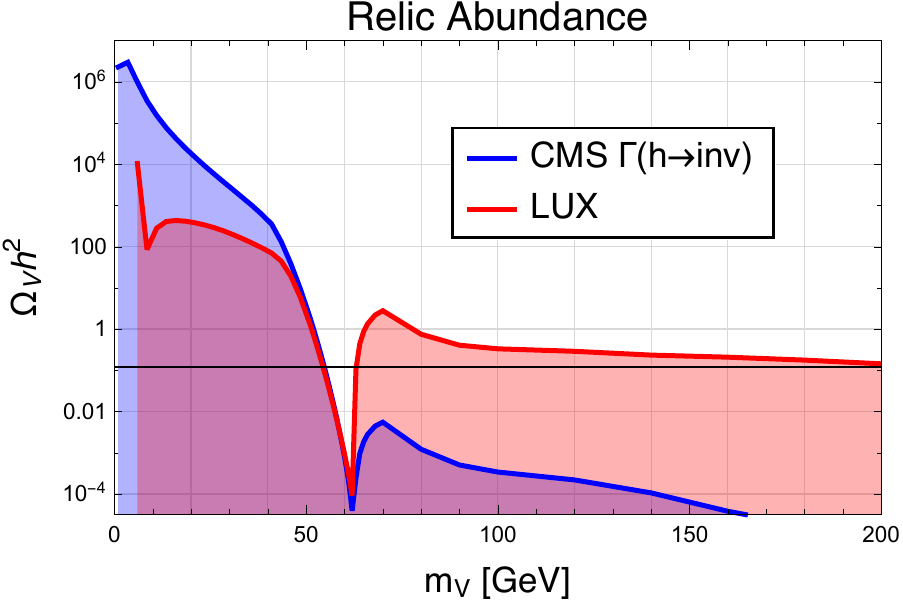}
	\caption{Left: Upper limits on $yg^2$ from VBF Higgs collider and direct detection constraints, with a fermion of mass 400 GeV. 
	Right: The corresponding lower limit on the relic abundance for a standard cosmology.}
\label{fig:simplim}
\end{figure}

Fig.~\ref{fig:simplim}, shows the collider and direct detection limits, plotted as the upper bound on $yg^2$ as a function of the dark matter mass, and the translation of
those upper limits into a lower limit on the relic abundance, assuming a standard cosmology, for the case when the single relevant fermion has a
mass of 400~GeV.  Despite the fact that the limits on the couplings are relatively weak, the conclusion is nonetheless that aside from a narrow region in the Higgs
funnel region, additional interactions would be required to deplete the dark matter relic density enough to saturate the observed relic density.

\subsection{Single Fermion with Scalar Mixing}
\label{sec:singleHP}

Building on the single fermion limit, we now allow for substantial $\lambda_P$ such that the radial modes of $H$ and $\Phi$ experience significant mixing, resulting in
two CP even scalars we denote by $h$ and $h_2$.  Describing this limit requires three additional free parameters, 
which we take to be the mass of the second scalar $m_{h_2}$,  $\langle \Phi \rangle = v_\phi$, and the Higgs-scalar mixing angle $\alpha$.

For small $\alpha$, the form factors of Eqn.~(\ref{eq:eftloop}) are shifted:
\begin{equation}
\begin{aligned}
A(p^2) &\rightarrow \left(1-\frac{\alpha^2}{2} \right) A(p^2) \\
B(p^2) &\rightarrow \left(1-\frac{\alpha^2}{2} \right) B(p^2)  - 2 \alpha \frac{m_V^2}{v_\phi}
\end{aligned}
\end{equation}
where the additional contribution is the tree level contribution to $B(p^2)$ from the induced $\Phi$ component in $h$. 
In addition to the shift in the effective $h$-$V$-$V$ coupling, the $h_2$ state acquires a coupling to the SM
given by the corresponding SM Higgs coupling multiplied by $\alpha$.

\begin{figure}
	\includegraphics[width=0.495\textwidth]{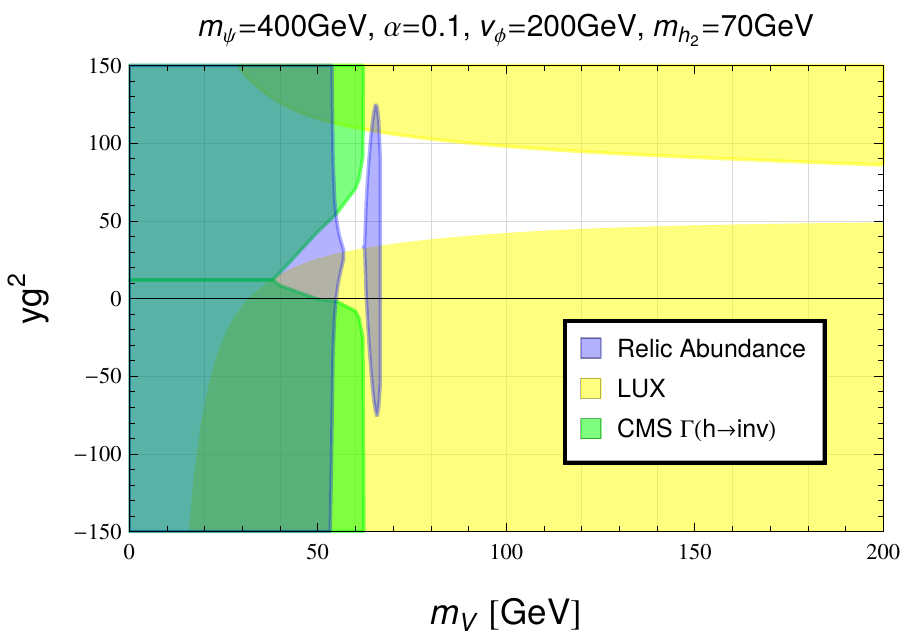}	
	\includegraphics[width=0.495\textwidth]{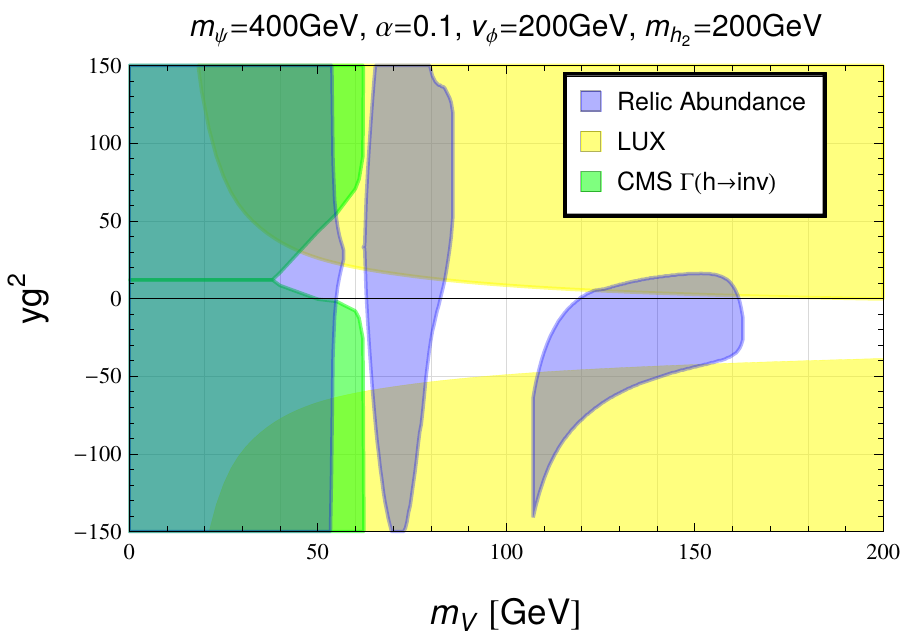}\\[0.65cm]
	\includegraphics[width=0.495\textwidth]{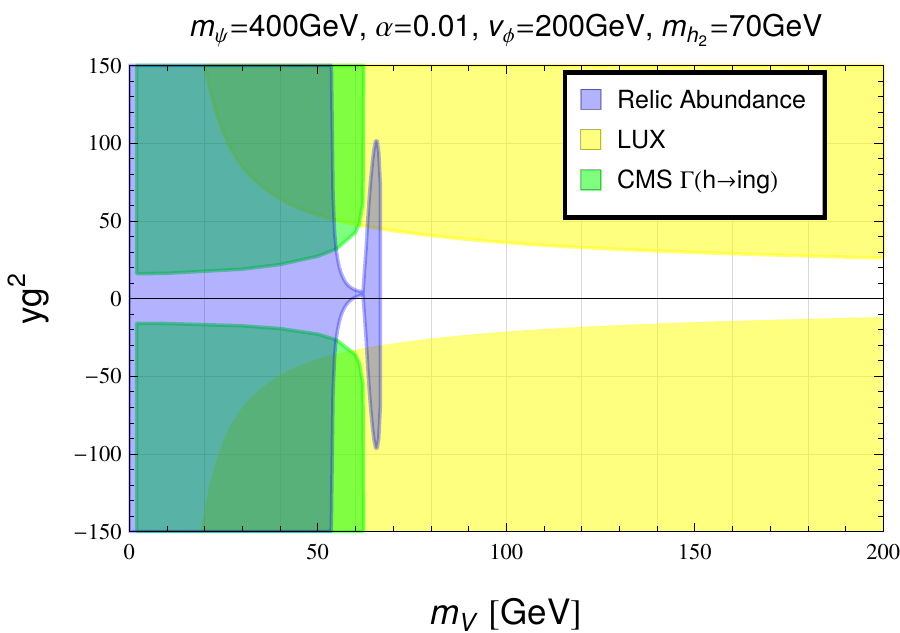}	
	\includegraphics[width=0.495\textwidth]{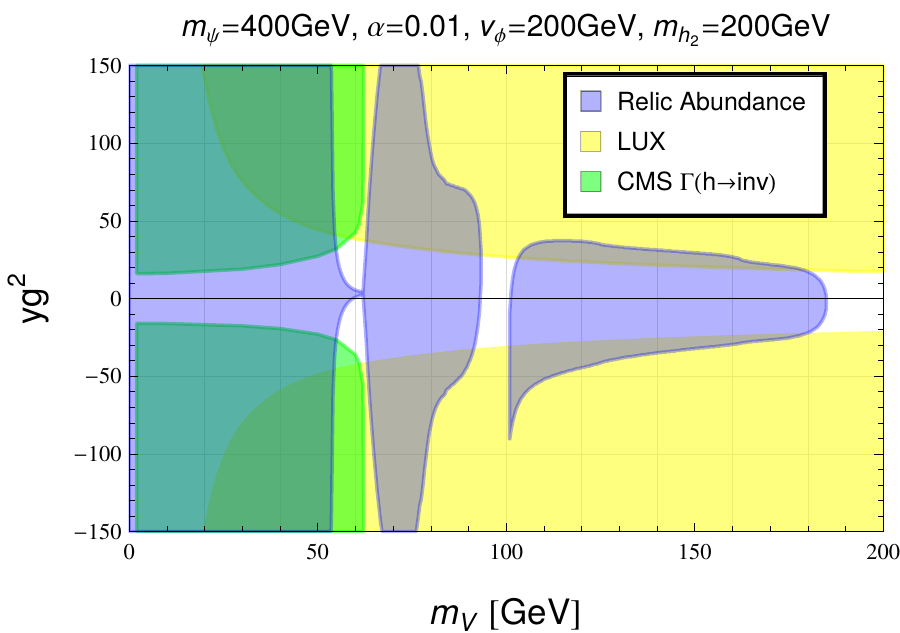}
	\caption{Exclusion regions on $yg^2$ for various parameters in the Higgs-Scalar mixing model. The left (right) two plots are for a scalar lighter (heavier) than the Higgs. The top (bottom) two plots are for a mixing angle of $\alpha = 0.1(0.01)$.}
\label{fig:wmixing}
\end{figure}

In Fig.~\ref{fig:wmixing}, we indicate the bounds on $y g^2$ as a function of the vector mass for various benchmark values of the remaining free parameters as indicated,
with shaded regions showing points excluded by the CMS invisible Higgs width bounds (green), and the LUX bounds on $\sigma_{\rm SI}$ (yellow).  Note the appearance
of ``blind spots" in the direct detection plane coming from interference between loop- and tree-level contributions to the $h$-$V$-$V$ vertex and/or
between $h$ and $h_2$ exchange \cite{Baek:2014jga}.
Blue shading
indicates regions where the dark matter is over-abundant in a standard cosmology.  Unshaded regions are allowed by current data and do not over-close the
Universe, with points close to the boundaries of the blue shading typically predicting a relic density close to the observed value.  Such regions consistent with
collider and direct searches are again typically in funnel regions for annihilation through $h$ and $h_2$, when it is heavier than $h$ itself.  Additional parameter space
also opens up for larger DM masses, where annihilation $VV \rightarrow h ~ h_2$ becomes viable.

\subsection{Full Matter Content}
\label{sec:fullF}

As our final limit, we return to $\lambda_P \ll 1$ but allow for all of the fermions to have comparable masses.  We consider three
benchmark sets of masses and Yukawa interactions summarized in Table~\ref{tab:bench}, which contains the model parameters associated with the fermion sector,
$m$, $m_n$, $y_\psi$, and $y_\chi$, as well as the resulting spectrum of neutral state masses $M_N$
and the coefficient of the $h$-$\bar{N}_i$-$N_j$ coupling in the mass basis, $Y_{ij}$, with the mass eigenstates ordered as $M_{N_1}>M_{N_2}>M_{N_3}$. Table~\ref{tab:benchgauge} and Eqn.~\ref{eq:gauge}, summarize the corresponding interaction of the gauge bosons with the new fermions.  With these quantities fixed, we explore the plane of the U(1)$^\prime$
gauge coupling $g$ and the mass of the dark matter $m_V$.

The new electrically charged fermions may be pair produced or produced in association with a new neutral fermion at colliders. For the regime of interest, the charged fermions decay solely to one of the neutral fermions and a W boson. The charged states are sufficiently similar to charginos in the MSSM that chargino searches may be applied. LEP searches require the charged fermion to be heavier than 100 GeV \cite{Heister:2002mn,Abdallah:2003gv,Acciarri:2000wy,Abbiendi:2002vz}. LHC searches find similar bounds which strengthen as the charged state becomes very long lived \cite{Aad:2014vma,Khachatryan:2014mma}. The lightest charged state among our benchmarks is 300 GeV, which is safe from these constraints.

Some couplings are taken to be quite large to help highlight the features of this model in observables. In choosing such large values for the gauge and yukawa couplings, one may be concerned that perturbativity breaks down or that higher order corrections should not be ignored. The latter case may even reduce the relic abundance when properly taken into account, which would open up available parameter space. Alternately, smaller couplings may be chosen which would reduce the range of viable dark matter masses. However, neither case appreciably alter our conclusions.

\begin{table}
\centering
\caption{Benchmark parameter sets, and resulting neutral fermion masses and Higgs couplings.}
\begin{tabular}{cccc||cc}
  $m$~~~ & $m_n$ & $~~~y_\psi~~~$ & ~~~$~y_\chi~$~~~ & ~~~~~$M_N$ (GeV)~~~ & ~~~Y \\
  \hline & & & & & \\
  800 GeV~~~~~ & 250 GeV & 1 & $-0.5$  & $\begin{bmatrix} 832 \\ 807 \\ 274 \end{bmatrix}$ & $\begin{bmatrix} -0.25 & -0.04 & 0.71 \\ 0.04 & -0.06 & 0.26 \\ -0.71 & 0.26 & -0.19 \end{bmatrix}$ 
  \\ & & & & & \\
  300 GeV~~~~~ & 200 GeV & 4 & $-2$ & $\begin{bmatrix} 848 \\ 810 \\ 238 \end{bmatrix}$ & $\begin{bmatrix} -3.0 & -0.81 & -0.56 \\ 0.81 & -3.0 & -0.47 \\ 0.56 & -0.47 & -0.02 \end{bmatrix}$
    \\ & & & & & \\
  500 GeV~~~~~ & 1000 GeV & 4 &~ 4 & $\begin{bmatrix} 1770 \\ 500 \\ 265 \end{bmatrix}$ & $\begin{bmatrix} -3.9 & 0 & 0.98 \\ 0~ & ~~~0~~~ & ~0 \\ -0.98 & 0 & -3.9 \end{bmatrix}$
\end{tabular}
\label{tab:bench}
\end{table}

\begin{eqnarray}
\label{eq:gauge}
\mathcal L_{gauge} &= e \left(\bar{E_1}\gamma^\mu E_1 - \bar{E_2}\gamma^\mu E_2\right)\left( A_\mu + \frac{(1 - 2s_w^2)}{2 c_w s_w} Z_\mu\right) ~ + ~ \frac{e}{2 c_w s_w} \bar{N_i}\gamma^\mu ~G^Z_{ij}~ N_j ~Z_\mu \nonumber\\ &+ \frac{e}{\sqrt{2}s_w} \left[ \left( \bar{E_1}\gamma^\mu ~G^{W1}_i~ N_i + \bar{N_i}\gamma^\mu ~G^{W2}_i~ E_2 \right) W_\mu^+ + h.c.\right] \nonumber \\ &+ g \left(\bar{E_i}\gamma^\mu E_i + \bar{N_i}\gamma^\mu N_i\right) V_\mu
\end{eqnarray}

\begin{table}
\centering
\caption{Gauge couplings matrices defined in Eqn.~\ref{eq:gauge} represented in fermion mass basis.}
\begin{tabular}{cccc||ccc}
  $m$~ & $m_n$ & $~~y_\psi~~$ & ~~$~y_\chi~$~~ & ~$G^Z$~ & ~~$G^{W1}$ & ~~$G^{W2}~$\\
  \hline & & & & & & \\
  800 GeV~~ & 250 GeV & 1 & $-0.5$ & $\begin{bmatrix} 0.01~\gamma^5 & -0.98 & -0.11 \\ -0.98 & 0.03~\gamma^5 & -0.17~\gamma^5 \\ -0.11 & -0.17~\gamma^5 & -0.04~\gamma^5 \end{bmatrix}$ & $\begin{bmatrix} 0.70 \\ 0.70-0.01~\gamma^5 \\ 0.08+0.12~\gamma^5 \end{bmatrix}$ & $\begin{bmatrix} 0.70 \\ -0.70-0.01~\gamma^5 \\ -0.08+0.12~\gamma^5 \end{bmatrix}$
  \\ & & & & & & \\
  300 GeV~~ & 200 GeV & 4 & $-2$ & $\begin{bmatrix} 0.20~\gamma^5 & -0.36 & 0.69 \\ -0.36 & 0.38~\gamma^5 & -0.35~\gamma^5 \\ 0.69 & -0.35~\gamma^5 & -0.59~\gamma^5 \end{bmatrix}$ & $\begin{bmatrix} -0.56+0.09~\gamma^5 \\ -0.26+0.36~\gamma^5 \\ 0.65+0.23~\gamma^5 \end{bmatrix}$ & $\begin{bmatrix} -0.56-0.09~\gamma^5 \\ 0.26+0.36~\gamma^5 \\ -0.65+0.23~\gamma^5 \end{bmatrix}$
    \\ & & & & & & \\
  500 GeV~~ & 1000 GeV & 4 & 4 & $\begin{bmatrix} 0 & -0.61 & 0 \\ -0.61 & 0 & 0.79~\gamma^5 \\ 0 & 0.79~\gamma^5 & 0 \end{bmatrix}$ & $\begin{bmatrix} -0.43 \\ -0.71 \\ 0.56~\gamma^5 \end{bmatrix}$ & $\begin{bmatrix} 0.43 \\ -0.71 \\ -0.56~\gamma^5 \end{bmatrix}$
\end{tabular}
\label{tab:benchgauge}
\end{table}

\begin{figure}
	\includegraphics[width=0.9\textwidth]{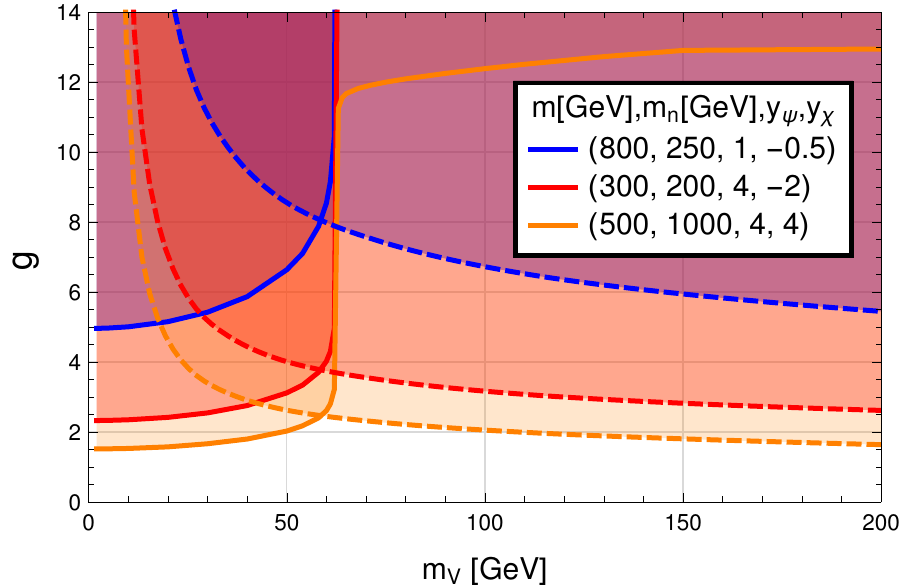}
	\caption{Upper bound on the gauge coupling, $g$, for the three benchmark parameters. VBF Higgs collider constraints are in solid and direct detection constraints are dashed lines.  Note that for the direct detection constraints we assume the local abundance of DM is $0.3$ GeV/cm$^3$ whereas the prediction from the model, for conventional thermal history, is often smaller, see Figure~\ref{fig:fullrelic}.}
\label{fig:fulllim}
\end{figure}

In Fig.~\ref{fig:fulllim}, we show upper bounds on $g$ as a function of the vector mass. We find that the collider and direct detection constraints are relatively weak,
often less constraining than perturbativity. 
Despite the mass of the lightest neutral state being similar for all three benchmarks, constraints are significantly stronger for the second and third cases, where the
Yukawa couplings are stronger.  In terms of the dominant contribution to the effective $h$-$V$-$V$ coupling, in the first and third models, the lightest neutral
state is the dominant contribution, whereas in the second benchmark model the lightest state has a small Yukawa coupling and is less important than the second lightest
state, which has a much larger coupling.

\begin{figure}
	\includegraphics[width=0.9\textwidth]{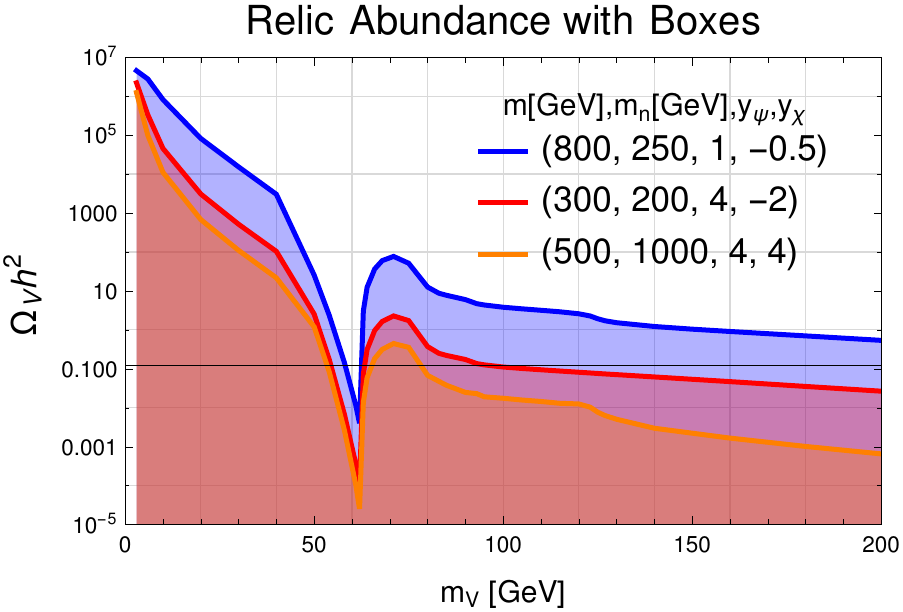}
	\caption{The vector relic abundance for the three benchmark parameters. The gauge coupling here is chosen to be $g=3.5$.}
\label{fig:fullrelic}
\end{figure}

In Fig.~\ref{fig:fullrelic}, we plot the relic abundance for the benchmark parameters with a large, fixed gauge coupling of $g = 3.5$,
to make comparisons between the benchmarks more apparent.  Note that for our second and third benchmark models, this value
is mildly excluded by limits on the invisible width of the Higgs for $m_V \leq 60$~GeV.
All benchmarks can be thermal relics when the vector can resonantly annihilate through a Higgs, causing the sharp dip at $m_V\sim m_h/2$. 
We also find that the second benchmark can attain a thermal relic for vector masses above 100 GeV, and third may be a thermal relic above 80 GeV. 
The success at larger DM masses is due to annihilation channels with two bosons in the final state.
Of the three benchmarks, the second has the lightest charged states. This allows efficient annihilation through loops involving the charged fermions, 
such as those which result in the $WW$ and $ZZ$ final states. The third benchmark, also benefits from this with slightly heavier charged states. 
However, this case also has large Yukawas causing a marked drop in the relic abundance when DM is heavy enough to annihilate to two Higgs bosons.

\section{Conclusion}
\label{conclusion}

We have explored a simplified model in which the dark matter is a spin one vector particle which interacts with the Standard Model predominantly through Higgs
exchange.  Unlike the more usually considered Higgs portal based on the quartic interaction $\lambda_P$, we mediate the interaction radiatively, via a loop
of heavy fermions charged under both the dark U(1)$^\prime$ as well as the SM electroweak interaction.  By construction, the theory is anomaly free, has
a heavy vector particle which is effectively stable, and leads to no large deviations in the properties of the SM Higgs.  This last feature, together
with the possibility to completely decouple the U(1)$^\prime$-breaking Higgs $\Phi$ from the SM are the primary
features which distinguish the radiative model from the quartic-induced Higgs portal as far as dark matter phenomenology is concerned.

Of course, the UV structure of the radiative model is also far richer, with a family of electroweakly charged particles whose decays produce gauge bosons
and missing momentum, a signature already under study in the context of the neutralinos and charginos of a supersymmetric theory.  These states are the
true avatars of the radiative Higgs portal.  The thermal relic density suggests that their masses are at most around TeV, raising the hope that they could
be found at the LHC run II or a future high energy collider.

\acknowledgments
AD is supported by the Fermilab Graduate Student Research Program in Theoretical Physics and in part by NSF Grant No.~PHY-1316792.
Fermilab is operated by Fermi Research Alliance, LLC under Contract No. DE-AC02-07CH11359 with the United States Department of Energy.
The work of TMPT is supported in part by NSF grant PHY-1316792 and by the University of California, Irvine through a Chancellor's Fellowship. 

\appendix
\section{$h$-$V$-$V$ Effective Vertex at One Loop}
\label{app:tri}

Here we outline the details of the triangle loop calculation.
The following results are for a single fermion species running in the loop. While the Higgs has off-diagonal couplings with the three neutral fermions in 
the mass basis, the vector only has diagonal couplings and thus only the diagonal Higgs interactions appear in the triangle diagrams. 
As a result, the functions $A$ and $B$ of Eq.~(\ref{eq:eftloop}) are the sum of the contributions from each individual fermion species.

Momenta are defined as in Fig.~\ref{fig:loop}, with $k_1$ and $k_2$ the two (on-shell) vector momenta coming into the diagram, and $p = - (k_1 + k_2)$ the momentum
incoming through the Higgs line.  In addition to the diagram shown explicitly in Fig.~\ref{fig:loop}, there is a second contribution related to it by
$k_1\!\leftrightarrow \! k_2$, $\mu \leftrightarrow \nu$.

The contribution to the matrix element from a single fermion of mass $m$ and Yukawa coupling $y$ is given by:
\begin{equation}
\mathcal M = g^2 \frac{y}{\sqrt{2}} \frac{i \pi^2 8m}{(2\pi)^4} \times \mathcal I^{\mu\nu}\left( k_1, k_2 \right) \times \epsilon_{\mu}(k_1) \epsilon_{\nu}(k_2)
\end{equation}
where,
\begin{equation}
\hspace*{-0.25cm}
\mathcal I^{\mu\nu}\left( k_1, k_2 \right) = 
\frac{1}{8m} \int \frac{d^d k}{i\pi^2}\frac{{\rm Tr}\left[(\slashed{k}+m)\gamma^{\nu}(\slashed{k}+\slashed{k}_2+m)(\slashed{k}-\slashed{k}_1+m)\gamma^{\mu} \right]}
{(k^2-m^2)((k-k_1)^2-m^2)((k+k_2)^2-m^2)}
 +(k_1,\mu\leftrightarrow k_2,\nu).
\end{equation}
Evaluating the trace in the numerator and making use of the fact that $k_1 \cdot \epsilon(k_1) = k_2 \cdot \epsilon(k_2) = 0$ for on-shell vectors results in,
\begin{equation}
{\rm Tr}[...] = 4m \left(g^{\mu\nu}(m^2-k_1\!\cdot \! k_2-k^2) + 4 k^{\mu}k^{\nu} + k_1^{\nu} k_2^{\mu}\right)~.
\end{equation}
After Passarino--Veltman decomposition \cite{Passarino:1978jh} we find,
\begin{equation}
\begin{aligned}
\mathcal I^{\mu\nu}\left( k_1, k_2 \right)
&=  \Big\{g^{\mu\nu}\Big[(4-d)C_{00} + m^2 C_0+ k_1\! \cdot \! k_2 (2C_{12}-C_0) - m_V^2(C_{11}+C_{22})\Big] \\ & 
~~~~~~~~~~~ + k_1^{\nu}k_2^{\mu}\Big[C_0-4C_{12}\Big] \Big\},
\label{eq:PVform}
\end{aligned}
\end{equation}
where the arguments of the $C$ functions are (uniformly) $C_0(k_1, k_2; m, m, m)$, etc.  

Reducing to scalar functions results in a finite expression of the form,
\begin{equation}
\mathcal I^{\mu\nu} = F_1(p^2, m)~ (k_1\! \cdot \! k_2 g^{\mu\nu} - k_1^{\nu}k_2^{\mu}) + F_2(p^2,m)~ g^{\mu\nu}
\end{equation}
corresponding to an effective three-point vertex described by
\begin{equation}
 - \left(\frac{g^2ym}{2\sqrt{2}\pi^2}\right) \left( \frac{1}{4} F_1(p^2, m)~ hV^{\mu\nu}V_{\mu\nu} + \frac{1}{2} F_2(p^2, m)~ h V^{\mu}V_{\mu} \right)
\end{equation}
where the form factors $F_1$ and $F_2$ are given by,
\begin{equation}
\begin{aligned}
F_1(p^2, m) &= \frac{1}{2bm^2(b-4a)^2} \Big\{2 m^2 (b-2a)\Big[4a(a-1)+b(1+6a-b)\Big] C_0  \\ & \qquad\qquad\qquad\qquad- 2a(2a+b) \Delta B_0 + (b-2a)(b-4a)\Big\} \\
F_2(p^2, m) &= \frac{4a^2}{b(b-4a)^2}\left\{2(b-a) \Delta B_0 - 2 m^2 \Big[4a(a-1)+b(1-2a+b)\Big] C_0 + 4a-b \right\}
\end{aligned}
\end{equation}
with,
\begin{equation}
a\equiv\frac{m_V^2}{4m^2}, ~~~~~b\equiv\frac{p^2}{4m^2}.
\end{equation}
The scalar integrals $C_0$ and $\Delta B_0$ can be expressed analytically as,
\begin{equation}
\begin{aligned}
C_0&=\frac{1}{4 m^2 b \beta} \sum_{j,k=1}^2 \left[2 {\rm Li}_2\left(\frac{1+(-1)^j \beta}{1+(-1)^k {\rm X} \beta}\right) - {\rm Li}_2\left(\frac{(1+(-1)^j \beta)^2}{1+(-1)^k2 {\rm Y} \beta+\beta^2} \right)  \right]~,\\
\Delta B_0 &\equiv B_0(m_V^2; m, m) - B_0(p^2; m, m)\\
 &= 2\sqrt{\frac{1-b}{b}} \arctan\left[\sqrt{\frac{b}{1-b}}\right]-2\sqrt{\frac{1-a}{a}} \arctan\left[\sqrt{\frac{a}{1-a}}\right].
\end{aligned}
\end{equation}
with
\begin{equation}
\beta \equiv\sqrt{1-4\frac{a}{b}}, ~~~{\rm X}\equiv\sqrt{1-\frac{1}{a}}, ~~~{\rm Y}\equiv\sqrt{1-\frac{1}{b}}~.
\end{equation}

As mentioned above, the coefficients $A$ and $B$ in Eq.~(\ref{eq:eftloop}) are given by the sum over the contributions from all three neutral
mediator fermions,
\begin{equation}
\begin{aligned}
 \quad A(p^2) &= \sum_i \left(\frac{g^2 y_i m_i}{2\sqrt{2}\pi^2}\right) F_1(p^2, m_i), \\
 \quad B(p^2) &= \sum_i \left(\frac{g^2 y_i m_i}{2\sqrt{2}\pi^2}\right) F_2(p^2, m_i).
\end{aligned}
\end{equation}

In the $m_V\rightarrow 0$ limit the two form factors become,
\bea
F_1 &=& \frac{1}{2b\, m^2}\left(1 + \frac{b-1}{2b} \left[{\rm Li}_2\left(\frac{2\sqrt{b}}{\sqrt{b}-\sqrt{b-1}}\right) + {\rm Li}_2\left(\frac{2\sqrt{b}}{\sqrt{b}+\sqrt{b-1}}\right) \right]\right) + \mathcal{O}(m_V^2)~,\\
F_2 &=& \frac{m_V^4}{4b^2m^4} \left( 4\sqrt{\frac{1-b}{b}} \arctan{\sqrt{\frac{b}{1-b}}} -5+\frac{1+b}{2b} \left[ {\rm Li}_2 \left( \frac{2\sqrt{b}}{\sqrt{b}-\sqrt{b-1}}\right) + {\rm Li}_2 \left( \frac{2\sqrt{b}}{\sqrt{b}+\sqrt{b-1}}\right) \right] \right) \nonumber \\
&& +\ \mathcal{O}(m_V^6)~.
\eea

\bibliography{VDM}

\end{document}